# Analysis of pedestrian behaviors through non-invasive Bluetooth monitoring

Yuji Yoshimura[a,b], Alexander Amini[a], Stanislav Sobolevsky[a,c], Josep Blat[b], Carlo Ratti[a]

[a] SENSEable City Laboratory, Massachusetts Institute of Technology, 77 Massachusetts Avenue, Cambridge, MA 02139, USA;

[b] Department of Information and Communication Technologies, Universitat Pompeu Fabra, Roc Boronat, 138, Tanger Building 08018 Barcelona, Spain;

[c] Center for Urban Science and Progress, New York University, 1 MetroTech Center, 19th Floor, Brooklyn, NY 11201, USA;

**Abstract** This paper analyzes pedestrians' behavioral patterns in the pedestrianized shopping environment in the historical center of Barcelona, Spain. We employ a Bluetooth detection technique to capture a large-scale dataset of pedestrians' behavior over a one-month period, including during a key sales period. We focused on comparing particular behaviors before, during, and after the discount sales by analyzing this large-scale dataset, which is different but complementary to the conventionally used small-scale samples. Our results uncover pedestrians actively exploring a wider area of the district during a discount period compared to weekdays, giving rise to strong underlying mobility patterns.

## Introduction

This paper analyzes pedestrians' mobility patterns during a special event, when their behaviors are believed to differ from those during a normal day. We focus on examining behavioral differences between discount days and other normal sale days in the shopping environment, in terms of their paths, consisting of the number of visited nodes, their sequential order, and the length of their stay in the district.

Tracking pedestrians' trajectories in a shopping area has long been identified as a critical research area for urban researchers. Scholars have pursued developing mathematical models of spatial shopping behaviors to predict the relevant impacts on the shopping environment, such as the number of visitors to shops or shopping centres of the city (Rasouli & Timmermans, 2013). Retailers also show their interests in such analysis due to the potential impacts on the turnover of their shops and the real estate values (Kurose, Borgers & Timmermans, 2001; Borgers & Timmermans, 2005; Borgers & Timmermans, 2014). Conversely, for the city authorities, such analysis can be used for the prediction of the consequences of urban planning and market shares (i.e., effects of store replacements), and for efficient crowd management (Schadschneider, Klingsch, Klupfel, Kretz, Rogsch & Seyfried, 2009).

In the framework of modeling and analyzing the dynamics of pedestrian behaviors in the shopping environment, research has been conducted to uncover the relationship between pedestrian behaviors and the features of the shopping environment (Borgers & Timmermans,1986a; Borgers & Timmermans, 1986b; Kurose, Borgers & Timmermans, 2001; Borgers & Timmermans, 2005; Dijkstra, Timmermans, de Vrie, 2009; Dijkstra, Timmermans & Jessurun, 2014; Kemperman, Borgers & Timmermans, 2009; Kurose, Deguchi & Zhao, 2009; Zhu & Timmermans, 2008; Borgers, Kemperman & Timmermans 2009). The data collection methodologies for those analyses are largely based on paper-and-pencil on-site interviews or questionnaires (see Borgers and Timmermans, 1986a, 1986b, for the detailed survey method). By limiting the survey area, the interviewers intercepted the pedestrians randomly and asked questions about the following: length of stay in the area, taken routes, number and order of visited shops, activity agenda regarding the number of planned and unplanned store visits, expenditures, purpose and motivation of the visit to the shopping district, and socio-demographic information. In order to obtain more accurate geo-localized information, state-of-the-art technologies are sometime employed. These technologies can help alleviate the shortcomings of traditional travel surveys (Rasouli & Timmermans, 2014; Shoval & Issacson, 2006; Stopher & Shen, 2011; Bricka et al., 2012). The comparative studies between GPS and travel survey data are discussed in Bricka & Bhat (2006), and a summary of modeling shopping destination choice is shown in Huang & Levinson (2015).

However, we identified several shortcomings of these methodologies: first, the application of the proposed methodology tends to be spatially limited to street segments or small shopping area, due to the considerable human efforts (i.e., direct observations). This makes it difficult to conduct comparative studies among districts or areas. Second, because of its labor intensity, the data collection can be performed only for a few hours or a few days, resulting, again, in the difficulty of comparative studies. Third, the proposed methodology largely excludes the factor of the walking conditions, such as the moving crowds in the shopping district (e.g., Borgers, Kemperman & Timmermans, 2009; Borgers & Timmermans, 2014).

The purpose of this paper is to compensate for the above-mentioned shortcomings. We propose the application of a Bluetooth detection technique (Eagle & Pentland, 2005; Kostakos, O'Neill, Penn, Roussos & Panadongonas, 2010; Delafontaine, Versichele, Neutens & van de Weghe, 2012; Nicolai, Yoneki, Behrens & Kenn, 2006; O'Neill, Kostakos, Kindberg, Sciek, Penn, Fraser & Jones, 2006; Versichele, Neutens, Delafontaine, van de Weghe, 2011; Yoshimura, Girardin, Carrascal, Ratti & Blat, 2012; Yoshimura, Sobolevsky, Ratti, Girardin, Carrascal, Blat & Sinatra, 2014) to monitor pedestrians' sequential movements throughout the shopping area. This technique enables us to generate a large-scale dataset of human mobility at the district scale because unannounced tracking methodologies make it possible to collect data for a longer period (i.e., one month), including weekends and special events.

We strategically placed one of our sensors in the metro station, which is one of the important entry/exit points of the study area. This indicates that the scope of our study is the analysis of the behaviors of pedestrians who come to and leave the district by the metro. This spatial selection corresponds with the previous research that intercepted shoppers randomly at the entry/exit point of the shopping district (e.g., Borgers & Timmermans, 2014). That is, our research method enables us to capture data about pedestrians' behaviors in the large-scale dataset in quantity, but without their attributions or inner thoughts; the previous research method captured data in small-scale quantity but with the attributions or inner thoughts. In addition, our dataset covers a longer period of time, which makes it possible to compare behaviors and how it changes depending on the time of day, and the day of the week. Thus, we try to examine "real and large-scale empirical data" to uncover the pedestrians' behavioral differences between sales periods and normal shopping days, in terms of the special trajectory, visited places, and length of stay in the determined district.

In the following section, we present our methodology and discuss how a Bluetooth detection technique can be a viable alternative to the existing approach, considering the large-scale datasets to be analyzed. In the third section, we describe the design of the experiment in the historical center of Barcelona and the features of the obtained datasets. In the fourth section, we analyze the pedestrians' behavioral differences between the sales period and normal shopping days. The paper concludes with a summary of findings and discussions.

**Methodology for the analysis framework**

The Bluetooth detection technique is based on the systematic observation method in the framework of the "unobtrusive measures" (Webb, Campbell, Schwartz & Sechrest, 1966). This makes use of people's unconsciously left *digital footprints* or "data exhaust" (Mayer-Schönberger & Cukier, 2013, page 113). The Bluetooth detection technique can be classified as a passive data collection technique, which is in contrast to the active data collection (see a review of data collection classification in Yoshimura et al., 2012, 2014). The following are the advantages of Bluetooth detection technique compared with other existing methods for the analysis of shopping behaviors.

First, this technique enables us to collect a large-scale set of pedestrian behaviors. Although interviews, questionnaires, direct observations (Flick, 2009), and active mobile phone tracking with or without GPS (Shoval, McKercher, Birenboim & Ng, 2013) can provide us with more accurate and detailed information about the actual trip made by a person, each of them tends to result in relatively small-scale sample sizes because they require asking participants to bring devices in advance. Second, passive data collection technique does not require the labor intensive (i.e., direct observation for several hours). This aspect enables us to collect the relevant dataset for a longer period, including during the weekend or a special event. Third, since Bluetooth detection is based on an unannounced tracking system, subjects are not aware of being tracked, resulting in unbiased behavioral data, as expected. Finally, the detection range of Bluetooth is much finer grained than the network-based passive mobile detection technique, which can also generate a large-scale dataset of human mobility (González, Hidalgo & Barabási, 2008). The detection range of the former is based on the antenna's coverage, which cannot identify pedestrians' locations between streets.

The unannounced tracking system aspect of the Bluetooth technique provides us the above-mentioned advantages over active data collection techniques. However, this aspect also may raise discussions about ethical and privacy issues. MAC address, which can be detected by Bluetooth, is not associated with any personal information (see Delafontaine et al., 2012, p.661), so the detection of such information from individual mobile devices is considered to be legal. Considering future use and its adaptation for the urban setting condition, including possible changes to the legislative framework in the future, we propose applying a hash algorithm (Stallings, 2011, p. 342-361) to our sensor in order to maintain the anonymity of the visitors' data by converting the MAC address into a unique pseudonym (Yoshimura et al., 2014). This point makes our research method different from other prior work, which is also based on all of the above-mentioned Bluetooth detection techniques.

**Study setting**

The 1$^{st}$ of February marks the start of the second major discount period in Barcelona, Spain and lasts until the end of the month. Thus, we selected a data collection period from 01/29/2009 to 02/20/2009, resulting in data from more than 4 million unique devices. The choice of this specific season for the research enabled us to analyze the impact of discount days on the pedestrians' behaviors compared with the other normal days.

*Study environment and study settings*

The supply of shops on each street is considered one of most important factors affecting pedestrian behaviors in a shopping environment (Borgers & Timmermans, 2005). The density of the population per retail shop in the study area is 0.072, which is the highest one among the district in Barcelona and the second highest one in terms of the dimension (i.e., density per retail shop). Our previous research uncovered that the number of transactions in this district is superior to the number of economic activities (Yoshimura et al., 2016).

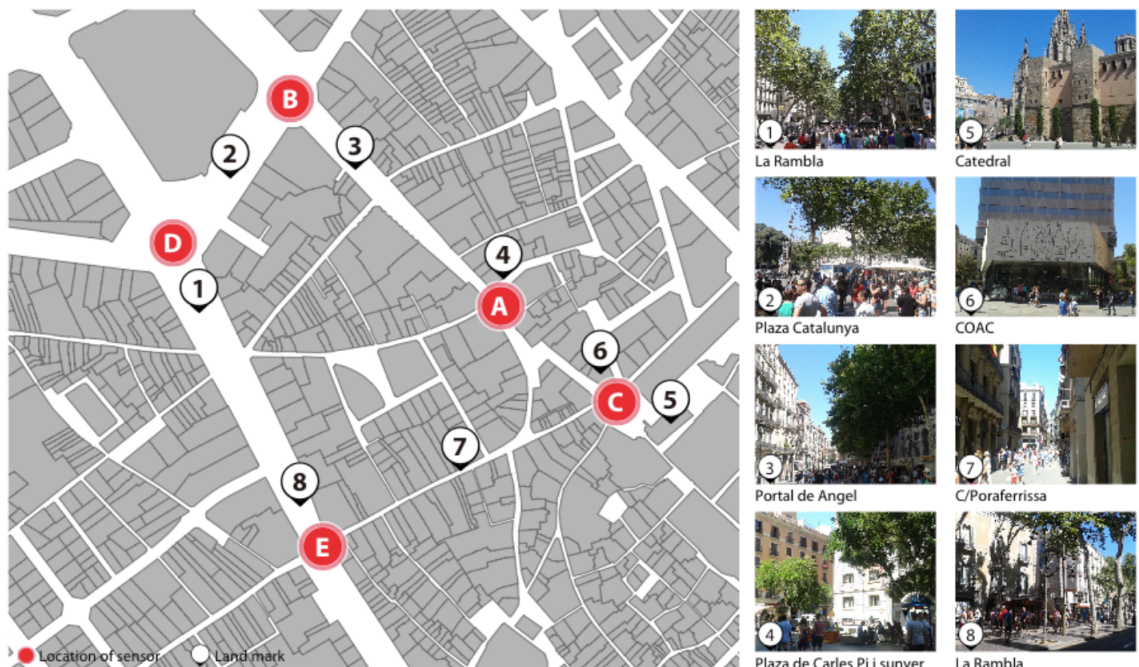

**Figure 1.** Location of five sensors (red circles) placed in the historical center of Barcelona. Key tourist attractions are numbered by white circles.

We deployed five Bluetooth sensors (nodes) for detecting the presence of pedestrians and their sequential movement between places in the historical center of Barcelona, Ciutat Vella (see Figure 1). The administrative and technical restrictions (e.g., protection against robbery and vandalism) largely determined the locations for their installation, and these restrictions sometimes prevented us from using the best locations for optimal detection and limited the number of sensors we installed. Although a higher number of sensors may provide more detailed information about the pedestrians' transitional movements, the five strategically selected places were enough to generate the most adequate and relevant information for our analysis. Those five selected points corresponded to locations in some of most congested and fluent pedestrian flows in the city of Barcelona. Sensor *C* was located in front of the Cathedral, which is one of the most important monuments in the city of Barcelona, attracting a large amount of tourists throughout the day. Sensor *E* was placed in the middle of La Rambla at Portaferrissa Street, which is the main street to enter Barri Gotic from La Rambla. Sensor *B* was placed at the beginning of Portal de Angel Street toward the Plaça de Carles Pi i Sunyer (sensor *A*), and the sensor *D* was placed in front of the metro station in La Rambla.

*Data Preparation*

The raw dataset included several errors and inadequate segments for our analysis. Firstly, it contained data derived from vehicles because node *B* and node *E* faced the traffic road, resulting in the collection of signals from vehicles. Also, there exists data from residents who live near the sensor locations. However, we can distinguish the signals from vehicles and pedestrians by looking to see if the path sequence contains node *D*. Node *D* is installed at the entrance and exit of the metro station where only pedestrians can pass. Thus, we filter them from the dataset. Also, we determined the residents who live near sensor locations by analyzing recurrences, such as devices staying constantly in a node during a whole day, and subsequently, removed these logs from the dataset.

After clearing the data, we identified subsets of the dataset, which started and finished at node *D*. *D-D* indicates pedestrians' length of stay in the district, resulting in over 100,000 data samples for the analysis (105,597).

**results**

In the following subsections, we present the results of our analysis, built around the previously described dataset. In section 4.1, we discuss the general statistical analysis of the dataset to compare the first Saturday (7$^{th}$ of February) with other Saturdays and weekdays, and in section 4.2 we present the most frequently appearing paths taken by pedestrians in each case.

*General statistical analysis of weekdays and Saturdays*

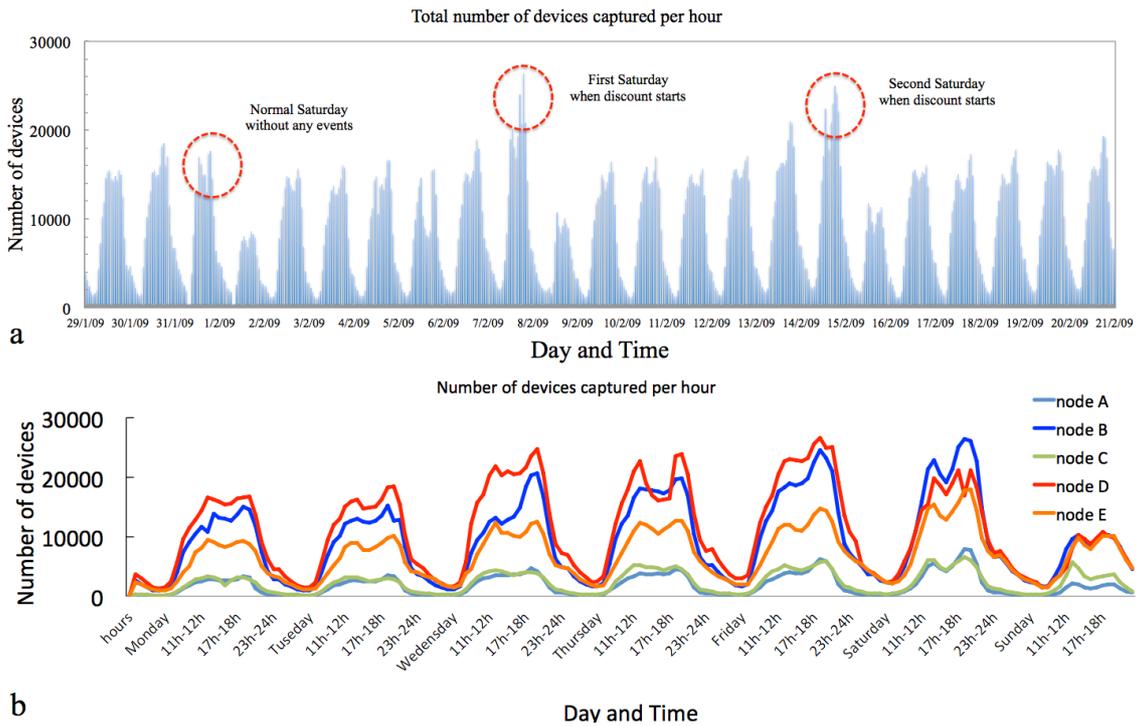

**Figure 2** (a) The number of devices captured per hour over the entire dataset. (b) Weekly patterns for the five captured areas during the study period.

Figure 2 (a) presents the number of devices captured per hour over the study period. We can see a consistent pattern of weekday activity and a weaker presence of traffic on Sundays, which amounts to almost half of a weekday. The traffic volume significantly increases on Saturday (7$^{th}$ and 14$^{th}$ of February). A graph of the weekly activity shows slightly different patterns of activity emerging in the five studied areas, with classic morning and afternoon peaks as well as differences between weekdays and weekends. In comparison to other areas, La Ramble (node E) didn't suffer from a large decrease of activity during the nighttime or on Sundays, probably due to the streams of tourist present in this area even when shops are closed. Figure 2 (b) reveals pedestrian patterns during the week, i.e., regular patterns and volumes during the weekdays with sharp decreases on Sundays. Node ***D*** always showed the highest volumes of traffic except on Saturdays, when node ***B*** overtook it with increased activity.

These divisional observations motivated us to divide all pedestrians into two groups in order to compare their behaviors: pedestrians traveling during the weekdays (Monday to Thursday) and those traveling on Saturdays. We excluded Fridays and Sundays from our analysis due to possible bias: it is clear that people behave differently at Friday when the weekend starts and Sunday when many of shops are closed. Thus, we compared the first Saturday (*F7*), when the discount started, with (1) the normal Saturday (*F31*) before the discount started, (2) the second Saturday (*F14*) after the discount began, and (3) weekdays (*W*). Thus, we determined the features of pedestrians' behaviors found particularly and distinctly on Saturdays.

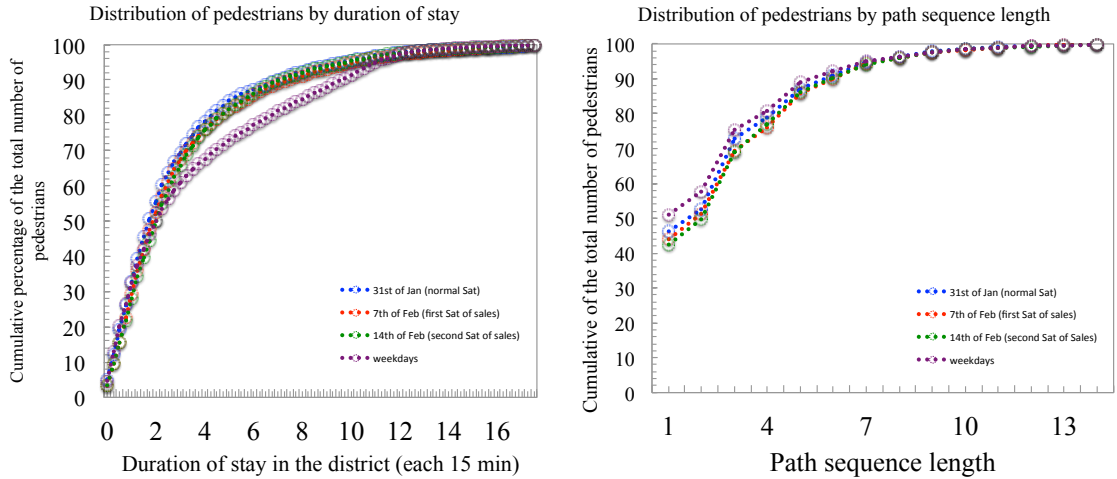

**Figure 3.** (a) The cumulative distribution of pedestrian's length of stay in the district. (b) The cumulative distribution of path sequence lengths.

Figure 3 (a) shows the cumulative distribution of the number of pedestrians against the length of their stay in the district aggregated in 15 minute bins on Saturdays and weekdays. We can observe that, while the behaviors during all Saturdays are quite similar, weekdays present a different distribution. This largely coincides with our intuition that pedestrians' behaviors might be different during weekdays and weekend. Pedestrians during Saturdays tend to rush to leave the district faster than pedestrians during the weekdays. Figure 3 (b) shows the cumulative distribution of the path sequence length during Saturdays and weekdays. Again, weekdays show a different distribution: otherwise, all of the rest present a similar distribution: pedestrians who visited only 1 node appear quite frequently, while pedestrians who visited 2 nodes do not show the expected growth rate. The average number of a path sequence length during weekdays is shorter than the corresponding average for Saturdays (i.e., 4.56 for $W$, 4.96 for $J31$, 5.11 for $F7$, 5.10 for $F14$). Although pedestrians on Saturdays tend to stay shorter in the district than on weekdays, they are likely to visit a larger number of nodes within the limited time than pedestrians during the weekdays.

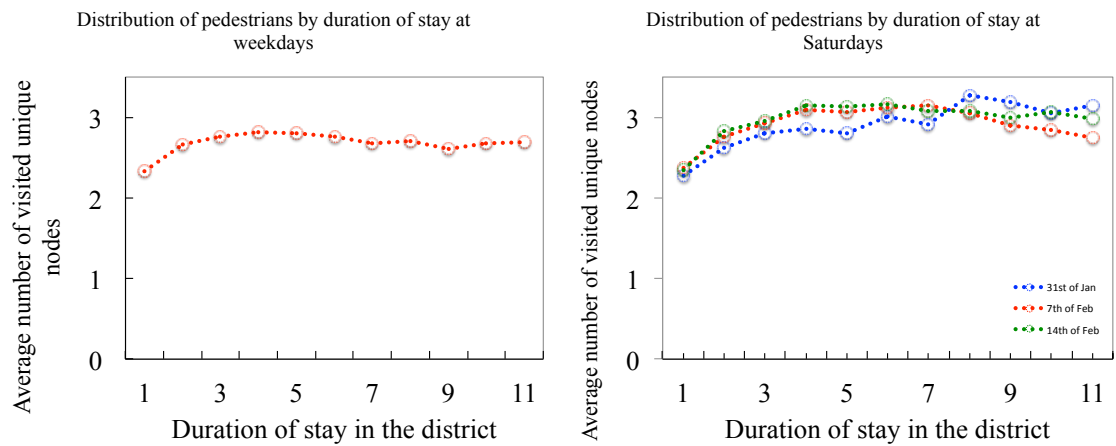

**Figure 4.** (a) The average number of unique visited nodes during weekdays. (b) The average number of unique visited nodes during Saturdays.

However, the path sequence length doesn't correlate to the area or the dimension visited, which the pedestrian explored. This is because they might just visit the same nodes multiple times, resulting in them circling only a limited area. Figure 4 (a) and (b) present the average number of unique visited nodes during weekdays and Saturdays respectively. While pedestrians during weekdays and *J31* (normal Saturday) tend to visit less than 3 nodes, those who visit during *F7* and *F14* surpass more than 3 nodes when they spend more than 3 hours in the area. This is the effect of node ***C***, which pedestrians rarely visit during weekdays and normal Saturdays, but much more frequently during discount Saturdays. Within Saturdays, the number of unique visited nodes during *J31* is always inferior to those during *F7* and *F14* when the pedestrians' length of stay is less than 7 hours.

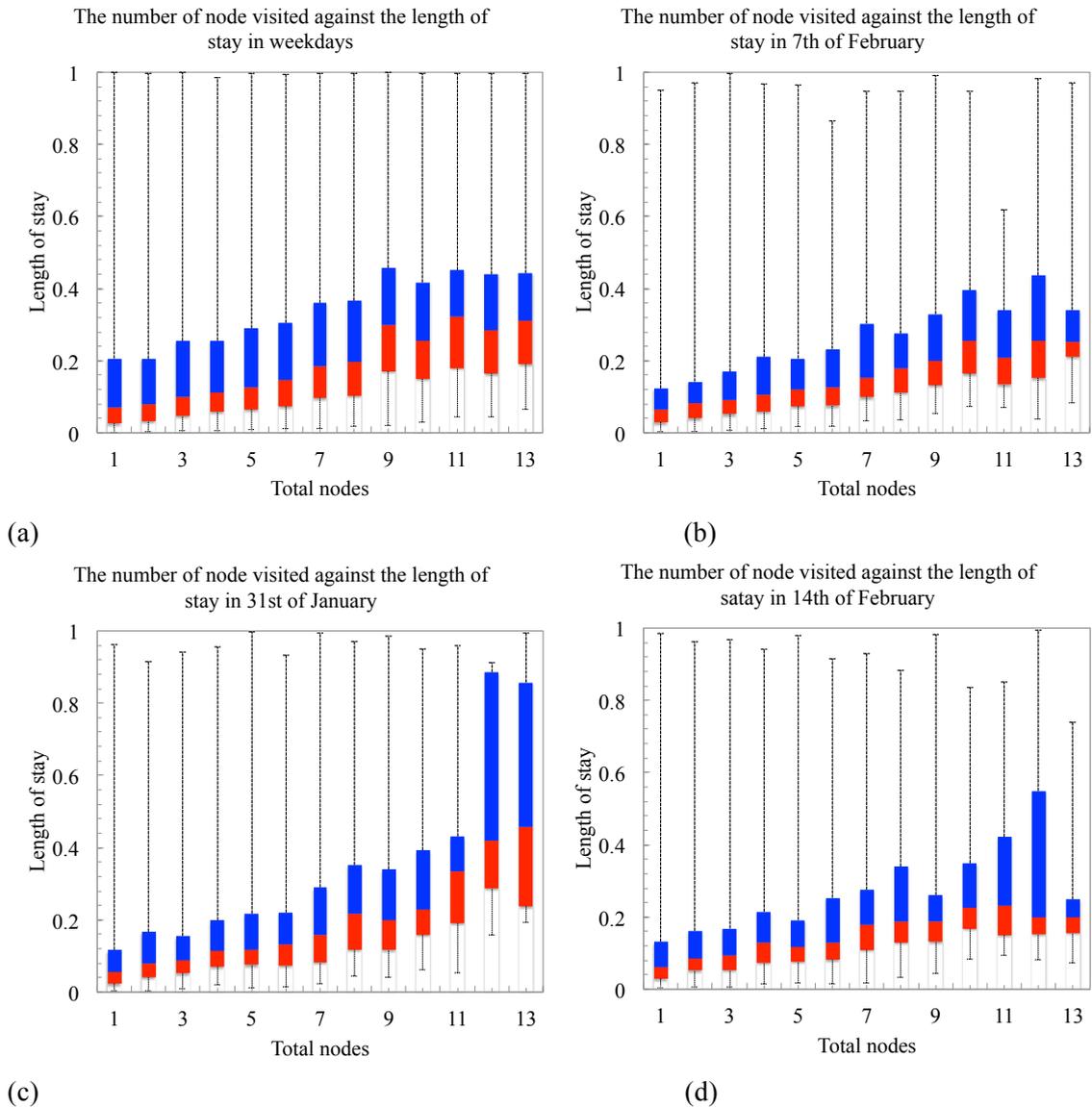

**Figure 5.** (a) The total number of visited nodes against the length of stay in the district during (a) *W* (b) *F7* (c) *J31* and (d) *F14*.

**Table 1.** The ρ and p-value of Spearman's rank correlation coefficient.

| | ρ | p-value |
|---|---|---|
| *Weekdays (W)* | | |

|  |  |
|---|---|
| | 0.2740 | 0 |
| 31$^{st}$ of January (J31) | | |
| | 0.4077 | 9.1164e-179 |
| 7$^{th}$ of February (F7) | | |
| | 0.3692 | 4.1179e-195 |
| 14$^{th}$ of February (F14) | | |
| | 0.3707 | 1.9305e-188 |

Figure 5 (a), (b), (c), and (d) show the relationship between the length of stay and the total number of visited nodes. We utilized a non-parametric correlation analysis (Spearman's rank correlation coefficient) because the variables in question do not follow a Gaussian distribution. We also included a series of boxplots to better explain the relationship between these variables. The correlation coefficient for the weekdays, *J31*, *F7*, and *F14* suggests a weak association between two variables ($\rho_w$ = 0.2740; $\rho_{J31}$ = 0.4077; $\rho_{F7}$ = 0.3692; $\rho_{F14}$ = 0.3707). In addition, the length of stay in the district during *F7* and *F14* tends to be shorter than the corresponding length during *W* and *J31*, independent from the number of visited nodes. This reveals, on the one hand, that pedestrians on *F7* and *F14* circulate more rapidly to a variety of different places and leave the district relatively quickly. This, in turn, creates a much higher turnover rate compared to the weekdays and *J31*. On the other hand, pedestrian behavior during *J31* (Saturday) is much more similar to the weekdays as opposed to Saturdays during the discount period.

All of these facts reveal that pedestrians during discounted Saturdays (*F7* and *F14*) actively explore different places rather than limiting their stay to a smaller area. In addition, they tend to spend shorter amounts of time in the district than pedestrians during weekdays and normal Saturdays, who tend to visit a larger dimension of the district.

We typically consider that a longer length of stay in the district may lead to an increase in the number of visited nodes, and vice versa. That is, the more time pedestrians spend in the district, the greater the possibility they have of exploring a larger number of nodes. However, our analysis revealed that there is no positive correlation between those two variables: rather our finding indicates a negative correlation which gets stronger during discount Saturdays compared to weekdays and normal Saturdays.

*Path patterns*

**Table 2.** Top 5 of the frequently appearing paths.

| Pedestrians whose length of path is more than 4 | less than or equal to 4 |
|---|---|
| *Weekdays* | |
|    D-B-D-B-D; 7.8% | D-B-D; 36.5% |
|    D-B-D-B-D-B-D; 2.0% | D-E-D; 13.6% |
|    D-B-A-B-D; 1.8% | D-B-E-D; 4.0% |
|    D-E-D-B-D; 1.4% | D-E-B-D; 0.7% |
|    D-B-D-E-D; 1.2% | D-A-B-D; 0.6% |
| *31$^{st}$ of January* | |
|    D-B-D-B-D; 10.4% | D-B-D; 33.8% |
|    D-B-D-B-D-B-D; 2.0% | D-E-D; 11.7% |
|    D-B-A-B-D; 2.0% | D-B-E-D; 3.6% |
| D-E-D-B-D; 1.5% | D-A-B-D; 0.8% |
| D-B-D-E-D; 1.3% | D-E-B-D; 0.6% |
| *7th of February* | |
| D-B-D-B-D; 6.5% | D-B-D; 28.6% |
| D-B-A-B-D; 2.1% | D-E-D; 14.6% |
| D-B-A-E-D; 2.0% | D-B-E-D; 3.9% |
| D-B-D-B-D-B-D; 1.7% | D-A-B-D; 0.9% |
| D-E-A-B-D; 1.6% | D-E-B-D; 0.6% |
| *14th of February* | |
| D-B-D-B-D; 6.8% | D-B-D; 26.9% |
| D-B-A-E-D; 2.3% | D-E-D; 14.8% |
| D-B-A-B-D; 1.9% | D-B-E-D; 3.7% |
| D-E-D-B-D; 1.8% | D-E-B-D; 0.9% |
| D-B-A-E-A-B-D; 1.5% | D-A-B-D; 0.7% |

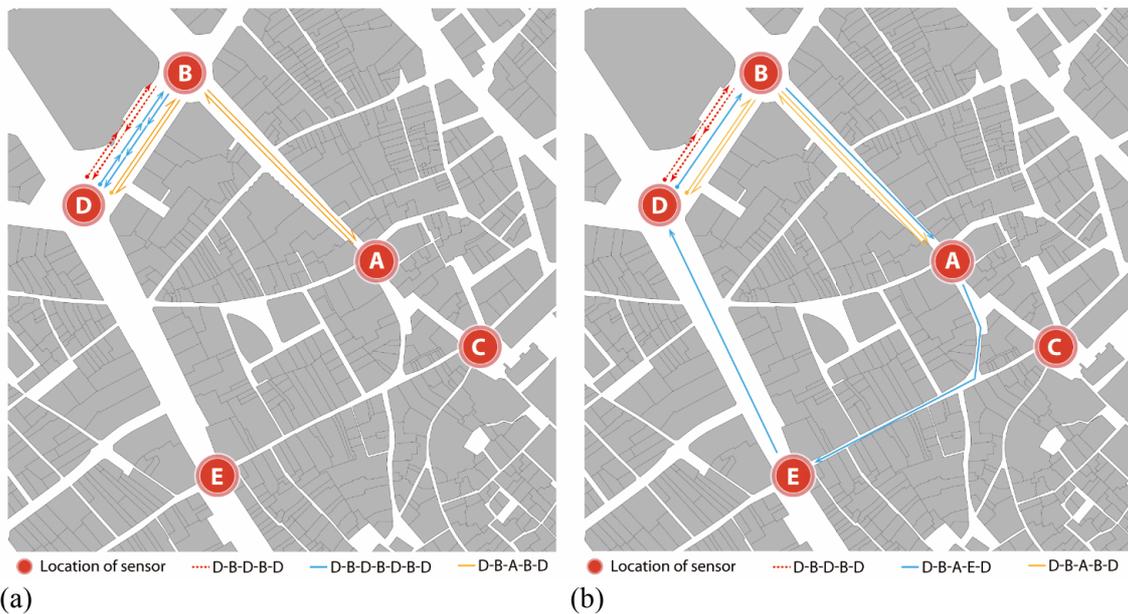

(a)                      (b)

**Figure 6.** (a) Visualization of the three most frequently appearing paths for *W* and *J31*. (b) The three most frequently appearing paths for *F7* and *F14*.

We measured the frequency of path appearance in each case and then normalized by the total number present in each group (Table 2). The most frequently appearing path for all groups (i.e., ***D-B-D-B-D***) indicates that pedestrians might explore La Rambla Street, but they rarely arrive in at Portaferrissa Street (node ***E***). Similarly, they are likely to walk down Portal de Angel Street but rarely arrive at Plaza de Carles i Sunyer (node ***A***) or COAC (node ***C***). Pedestrians tend to remain circulating in a limited area around the upper part of Ciutat Vella among the street along with Plaza Catalunya and the beginning of La Rambla. This tendency is stronger for pedestrians during a normal Saturday and weekdays compared to the first Saturday (*F7*) and second Saturday (*F14*) when the discount starts.

***D-B-A-E-D*** and the reverse sequence (i.e., ***D-E-A-B-D***) can be considered a key feature of pedestrians on the 7th of February. This path suggests that pedestrians visited Portaferrissa Street (node ***E***) through the middle of Portal de Angel Street (node ***A***) from the beginning of Portal de Angel Street (node *B*), then returned to the beginning of La Rambla (node *D*).

The path ***E-A*** or ***A-E*** is more frequently used by pedestrians during *F7* than during *W* and other Saturdays. We computed the probability for routes upon pedestrian arrival at node *A*. In the case of weekdays and *J31* (normal Saturday), pedestrians are more attracted to node *C* than node *E*. However, during discount periods, they are more attracted to node *E* than node *C*. This suggests that many more pedestrians during *F7* and *F14* select to approach La Rambla through Portaferrissa Street rather than via the Cathedral. On the first and second Saturday in February, pedestrians seem to be drawn more toward landscape consisting of small retail shops than the touristic perspectives of the Cathedral.

Conversely, most of pedestrians during the *J31, F14*, and *W* selected to approach to node *D* upon arriving at node *E* (i.e., 62.0%, 71.5%, and 68.1%): pedestrians during *F7* choose to move to node *A* (26.1%) and node *C* (23.0%). This indicates that during the first Saturday of the discount period, pedestrians used Portaferrissa Street much more than on weekdays and other Saturdays.

This tendency can be also observed when we focus on pedestrian transition time between the pairs of nodes. Pedestrians tended to move slowly when they traveled from Portaferrissa Street (node *E*) to Plaza de Carles Pi i Sunyer (node *A*), and this tendency was even stronger on the 7$^{th}$ of February compared to weekdays. Conversely, their visit duration becaome significantly longer when they traveled the opposite direction (i.e., from node *A* to node *E*), with almost a 40 minute difference in travel time. During weekdays, this difference became less significant (i.e., 13 minutes), although the trend was reversed, with the transition from *E* to *A* taking longer than the one from *A* to *E*.

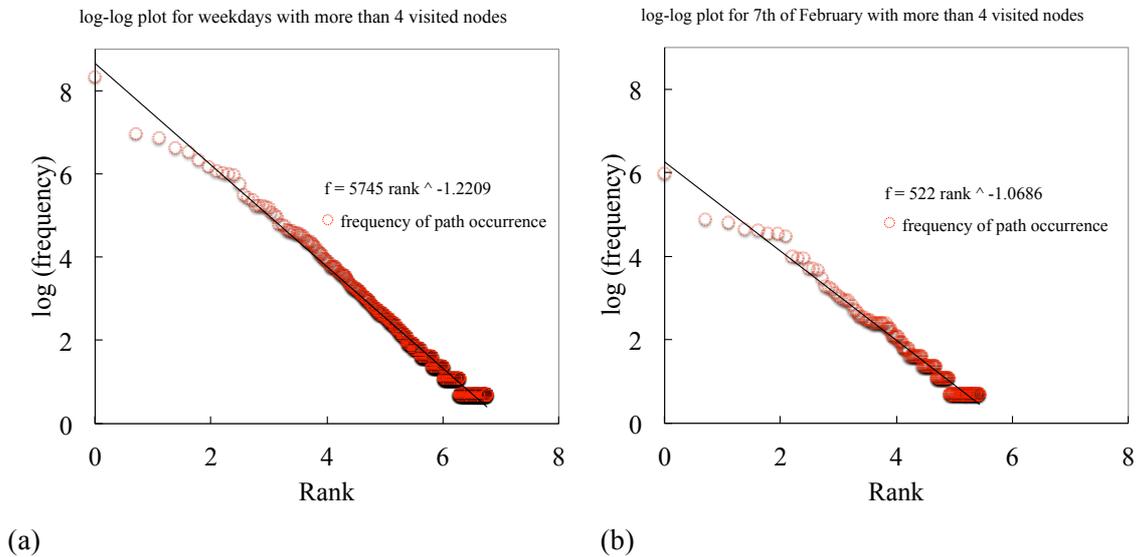

(a)　　　　　　　　　　　　　　　　　　(b)

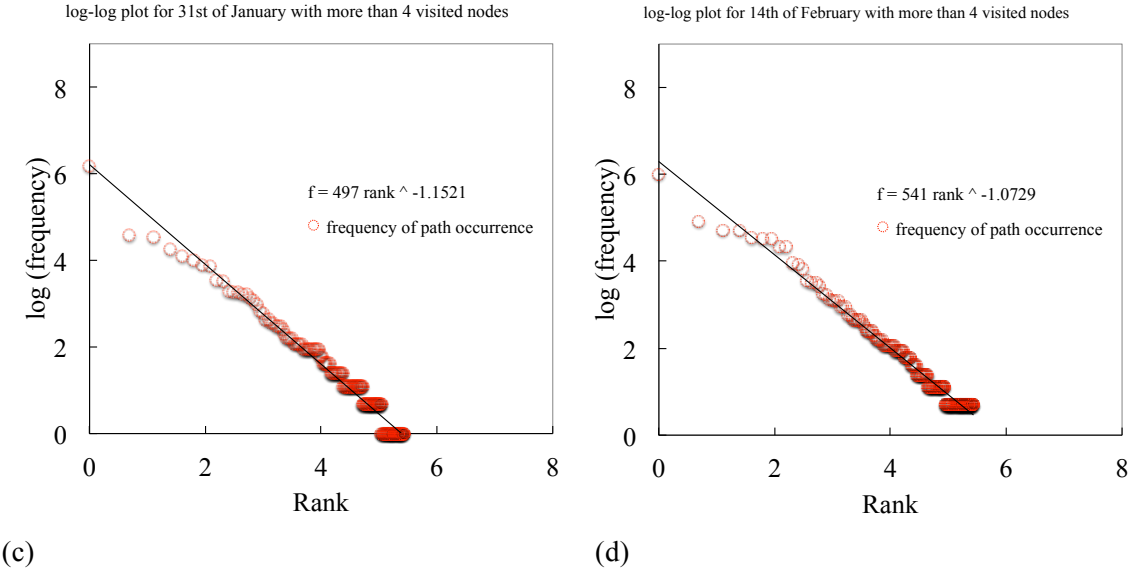

(c)                                        (d)

**Figure 7.** Rank distribution of pedestrians for (a) *W* (b) *F7* (c) *J31*, and (d) *F14* whose path length is more than 4.

**Table 3.** The slope of the line of best fit in log-log rank plot of path frequencies.

| Pedestrians whose length of path is more than 4 | less than or equal to 4 |
|---|---|
| *Weekdays* | |
| -1.2209 | -3.0163 |
| *31$^{st}$ of January* | |
| -1.1521 | -2.257 |
| *7$^{th}$ of February* | |
| -1.0686 | -2.8228 |
| *14$^{th}$ of February* | |
| -1.0729 | -2.0051 |

We started by examining the strength of pedestrian patterns through their respective paths. We visualized a log-log rank plot of the frequency of path for weekdays and Saturdays, whose path length is more than 4.

Pedestrians on *F7*, who visited 4 or more nodes, showed the strongest resemblance to a power law among all groups (slope = -1.0686). This is much stronger than the pedestrians who visited 4 or more nodes on weekdays (-1.2209) and other Saturdays (-1.1521 for *J31* and -1.0729 for *F14*). Conversely, the strength of patterns that visit less than 4 nodes is much lower, indicating that these paths have much greater variability in terms of type.

All of these facts indicate that pedestrians on the discount day (*F7*) tend to explore the district actively by visiting more places rather than staying in a relatively limited area. In addition, we found a correlation between an increased number of pedestrians visiting a given number of nodes and an increased tendency to use the same path sequence between said nodes. This tendency gets even stronger for the first discount day in February compared to weekdays and other Saturdays. Furthermore, these pedestrians visit these places over a shorter holistic timespan compared to pedestrians on the weekdays, with the exception of several intermediate streets (i.e., **A-E, A-B, B-D,** and

**D-E**). Conversely, the potential area that tends to be explored by pedestrians during the normal Saturday and weekdays is quite limited. Pedestrians generally remain in the upper part of Ciutat Vella without visiting the Cathedral, the middle of La Rambla, and Portal de Angel.

**conclusions**

This paper analyzed the differences between the behavioral patterns of pedestrians on discount days compared to normal days. We installed five Bluetooth sensors in the historical center of Barcelona and collected data over the course of one month. This systematic and unobtrusive observation method enabled us to obtain a very large-scale dataset of pedestrian mobility in terms of the number of visited nodes, the sequential order of their visits, and the length of stay in the district.

Results showed that pedestrians' behaviors on the Saturdays during the discount periods (*F7* and *14F*) are quite different from the normal Saturday (*J31*) and weekdays (*W*). We intuitively think that pedestrians behave differently during the weekend from weekdays, and our result supported this. However, our analysis uncovered that pedestrians' behaviors on normal Saturdays are more similar to their behaviors on weekdays than on Saturdays during the discount period.

During Saturdays in a discount period, pedestrians actively explored the district by visiting all nodes, including the Cathedral (node *C*), which is rarely visited by pedestrians on weekdays. We speculated that pedestrians on discount days might rush to visit the district as much as they can within their limited length of stay. However, we also discovered that these types of pedestrians tend to spend more time than corresponding ones on weekdays, depending on the streets they visited and the direction of transitions between the pair of nodes in said streets. Finally, we revealed that a pedestrian's sequential movement between nodes has underlying patterns in terms of the number of visited nodes and their order. This pattern is much stronger on discount Saturdays (*F7* and *F14*) than normal Saturday (*J31*) and weekdays, which creates a relief of the features of pedestrian behaviors during the discount days. By visualizing the number of pedestrians against the path type, we see the emergence of the power law distribution, indicating that most pedestrians use only a few path types, and most of path types are used only by a few pedestrians.

These in-depth mobility analyses in the shopping area were not possible prior to our study. Modeling and simulating pedestrian behaviors were frequently conducted in order to uncover underlying patterns in the shopping area (Borgers et al, 2009). However, they may be less appropriate to deal with them, considering the crowd effect in the longer-term, including special events, due to the data collection methodology employed for the research. Previous studies tend to use a small-scale sample set collected manually to estimate the parameters and calibrate the results of the simulations. Our approach and methodologies also have several advantages over the previously conducted research using state-of-the-art-technologies: the network-based passive mobile phone detection technique (Gonzales et al., 2009) cannot detect human movement at the street scale, and GPS based tracking (Shoval et al., 2013) and RFID-based studies (Larson et al., 2005) can only provide a small sample sets. The Bluetooth detection technique is frequently used to collect the pedestrians' sequential movement (Kostakos et al., 2010; Versichele et al., 2010; Delafontaine et al., 2012; Yoshimura et al., 2014) but is rarely applied in the shopping environment to examine the pedestrians' behavioral differences between discount and normal days. Thus, our analysis sheds a new light on unknown aspects of pedestrian behaviors in terms of the number of visited places, the order of their visits, and their length of stay in the district, thus making a significant contribution to the research of shopping behaviors.

Despite the above contributions, this method has some shortcomings, which resulted in the limitations of our research. First, a Bluetooth proximity sensor only knows the time-stamped sequence of individual transitions between nodes (i.e., sequence of *A-B-D*) for a given mobile device, resulting in the impossibility of determining the actual paths of activities between consecutive detections. Second, this study does not aim at revealing a consumer's decision-making process or their value consciousness because our dataset does not capture their inner thoughts or other subconscious information typically derived from interviews, questionnaires and participatory observation (Flick, 2009). Finally, our sampling contains some bias in terms of a person's attributions. Bluetooth detection enables us to detect the mobile device, given that said device has Bluetooth activated, indicating that the dataset consisted only of people who had Bluetooth on and activated on their devices. This required calculating the sample's representativeness and is typically conducted by using a short-term estimation via manual counting (Versichele et al., 2012). To counter this, Yoshimura et al., (2014) conducted a long-term comparison of ticket sales in a museum and a sensor installed in a same place, revealing 8.2% of visitors activated Bluetooth on their mobile phone.

Future progress on these issues will make it possible to validate the methodology and formulate more refined analytical framework.